\RequirePackage{fix-cm}
\documentclass[a4paper,10pt]{article}
\usepackage{datetime}
\usepackage{graphicx}
\usepackage{mathptmx} 
\usepackage[T1]{fontenc}
\usepackage[utf8]{inputenc}
\usepackage[hidelinks,colorlinks=true,linkcolor=blue,bookmarks=false]{hyperref}
\hypersetup{citecolor=blue, urlcolor=blue}
\usepackage[all]{hypcap}
\usepackage{xcolor}
\usepackage[nodisplayskipstretch]{setspace}
\usepackage{mathrsfs,amsmath}
\usepackage{amsbsy}
\usepackage{bbold}
\usepackage{caption}
\usepackage{color}

\title{A pore-scale model for permeable biofilm: numerical simulations and laboratory experiments}
\author{David Landa-Marb\'an${}^{1}$ \and Na Liu${}^2$ \and Iuliu Sorin Pop${}^{1,3}$ \and Kundan  Kumar${}^1$ \and Per Pettersson${}^2$ \and Gunhild Bødtker${}^2$ \and Tormod Skauge${}^2$ \and Florin A. Radu${}^1$}
\date{}
\begin{document}
\maketitle
\noindent ${}^1$ Department of Mathematics, Faculty of Mathematics and Natural Sciences, University of Bergen, All\'egaten 41, P.O. Box 7803, 5020 Bergen, Norway.\\[5pt]
${}^2$ Uni Research CIPR, P.O. Box 7800, N-5020, Bergen, Norway.\\[5pt]
${}^3$ Faculty of Sciences, Hasselt University, Campus Diepenbeek, Agoralaan building D, BE3590 Diepenbeek, Belgium.\\[5pt]
Corresponding author: David Landa-Marb\'an (E-mail: David.Marban@uib.no)
\begin{abstract}
In this paper we derive a pore-scale model for permeable biofilm formation in a two-dimensional pore. The pore is divided in two phases: water and biofilm. The biofilm is assumed to consist of four components: water, extracellular polymeric substances (EPS), active bacteria, and dead bacteria. The flow of water is modeled by the Stokes equation whereas a diffusion-convection equation is involved for the transport of nutrients. At the water/biofilm interface, nutrient transport and shear forces due to the water flux are considered. In the biofilm, the Brinkman equation for the water flow, transport of nutrients due to diffusion and convection, displacement of the biofilm components due to reproduction/dead of bacteria, and production of EPS are considered. A segregated finite element algorithm is used to solve the mathematical equations. Numerical simulations are performed based on experimentally determined parameters. The stress coefficient is fitted to the experimental data. To identify the critical model parameters, a sensitivity analysis is performed. The Sobol sensitivity indices of the input parameters are computed based on uniform perturbation by $\pm 10 \%$ of the nominal parameter values. The sensitivity analysis confirms that the variability or uncertainty in none of the parameters should be neglected.

\noindent\textbf{keywords}: Biofilm $\cdot$ Numerical simulations $\cdot$ Laboratory experiments $\cdot$ Microbial enhanced oil recovery $\cdot$ Porosity 
\end{abstract}
\newpage
\begin{tabular}{l l}
\textbf{List of Symbols}\\
$c$ & Nutrient concentration\\
$D$ & Nutrient diffusion coefficient\\
$d$ & Biofilm thickness\\
$J$ & Nutrient flux\\
$k$& Permeability\\
$k_{res}$ & Bacterial decay rate coefficient\\
$k_{str}$ & Stress coefficient\\
$k_{n}$ & Monod-half nutrient velocity coefficient\\
$L$ & Pore length\\
$p$ & Pressure\\
$q$ & Water velocity\\
$S$ & Tangential shear stress\\ 
$T$ & Time\\
$u$ & Velocity of the biomass\\
$U$ & Reference water velocity\\
$W$ & Pore width\\
$Y$ & Growth yield coefficient\\
\textbf{Greek Symbols}\\
$\mu$ & Dynamic viscosity\\
$\mu_n$ & Maximum rate of nutrient utilization\\
$\nu$ & Unitary normal vector\\
$\nu_n$ & Interface velocity\\
$\Phi$ & Growth velocity potential\\
$\rho$ & Density\\
$\tau$ & Unitary tangential vector\\
$\theta$ & Volume fraction\\
\textbf{Subscripts/superscripts}\\
$a$ & Active bacteria\\
$b$ & Biofilm\\
$d$ & Dead bacteria\\
$i$ & Input\\
$o$ & Output\\
$e$ & EPS\\
$w$ & Water\\
\textbf{Abbreviations}\\
ALE & Arbitrary Lagrangian Eulerian\\
EPS & Extracellular polymeric substance\\
MEOR & Microbial enhanced oil recovery\\
\end{tabular}
\section{Introduction}
A biofilm can be defined as an aggregation of bacteria, algae, fungi, and protozoa enclosed in a matrix consisting of a mixture of polymeric compounds, primarily polysaccharides, generally referred to as extracellular polymeric substance (EPS) (\cite{Vu:Article:2009}). Biofilms are present in many systems, with beneficial applications in some areas, for example in medicine, food industry, and water quality (\cite{Kokare:Article:2009}). In our research, we are interested in studying the biofilm to improve the oil extraction. Microbial enhanced oil recovery (MEOR) is an oil enhanced recovery method relying on microorganisms and their metabolic products to mobilize residual oil in a cost-effective and eco-friendly manner. The particular MEOR mechanism that we are concerned with in this work is called selective plugging. This mechanism consists of growing the bacteria in the high permeable zones in the reservoir and thereby clogging the preferential water flow paths. Consequently, the water will be forced to flow in new pores and more oil will be recovered. MEOR is not yet completely understood and there is a strong need for mathematical models to be used for improving these technologies.

The percentage of water in biofilms constitutes up to 97\% (\cite{Ahmad:Book:2017}). In the biofilm, cell clusters may be separated by interstitial voids and channels, which create a characteristic porous structure (\cite{Picioreanu:Article:2000}). The proportion of EPS in biofilms can comprise approximately 50-90$\%$ of the total organic matter (\cite{Donlan:Article:2002,Vu:Article:2009}). Flow velocity near a biofilm changes from a maximum in the bulk solution to zero at the bottom of the biofilm (\cite{Lewandowski:Book:2003}). In different biofilms, the mechanism of nutrient transport near the biofilm surface and within the biofilm can be dominated by convection or diffusion (\cite{Schwarzenbach:Book:1993}).  

Most of the biofilm models are based on simplifying assumptions, e.g. impermeability, a constant biofilm density, and accounting for diffusion but neglecting convection for transport of nutrients (\cite{Alpkvist:Article:2007,Duddu:Article:2009,Schulz:Article:2016}). Novel mathematical models must be built to improve accuracy and enhance confidence in numerical results. In \cite{Landa:Article:2017} we built a mathematical model for MEOR including the oil-water interfacial area. Pore-scale models are used to derive parameters and functional relationships for the core-scale models (\cite{Noorden:Article:2010,Ray:Article:2013,Bringedal:Article:2016,Bringedal:Article:2017}). In this work, we propose a pore-scale biofilm model including a permeable biofilm, a variable biofilm density, and transport of nutrients due to convection and diffusion.  

The resulting mathematical model involves coupled partial differential equations. Further, the biofilm-water interface location changes over time, and therefore is a free boundary problem. Numerical methods for solving free boundary problems are an active research field (\cite{Esmaili:Article:2017,Gallinato:Article:2017}). The arbitrary Lagrangian-Eulerian (ALE) method is used to track the position of the biofilm-water interface (\cite{Donea:Article:2004}). In biofilm and reactive flow modeling involving free boundary,  it is common to use decoupling techniques to find a numerical solution (\cite{Alpkvist:Article:2007,Peszynska:Article:2016,Kumar:SISC:2013}). In our case, a segregated finite element algorithm is used to solve the mathematical equations.

Due to the cost of performing laboratory experiments to accurately estimate material parameter values, it is of great interest to perform a sensitivity study with respect to the impact of a set of input parameters on certain model output quantities of interest. This ensures that critical parameters are identified. Moreover, for parameters scoring low in sensitivity estimates, less accurate parameter estimates can be justified. Global sensitivity analysis using Sobol indices is a means of quantifying the relative impact of a function of interest in terms of a set of varying input parameters (\cite{Sobol:Article:2001}). This is computationally prohibitive for problems with a large number of input parametes, but the computational cost can be significantly reduced by computing the Sobol indices using the generalized polynomial chaos framework (\cite{Xiu:Article:2002,Sudret:Article:2008}).

In this general context, the objective of the present article is to develop and implement an accurate numerical simulator for biofilm formation.

To summarize, the new contributions of this work are:
\begin{itemize}
\item the development of a multidimensional, comprehensive pore-scale mathematical model for biofilm formation,
\item the inclusion of a biofilm porosity,
\item the inclusion of nutrient transport inside the biofilm due to convection and diffusion, and
\item the calibration of the mathematical model with the laboratory experiments. 
\end{itemize}

We emphasize that the model development here is performed in close relationship with the physical experimental observations. In particular, this is the first model that takes into account the porous structure of biofilm. It is through the experiments that we identify the key processes and variables that need to be considered. Accordingly, we compute some of the parameters (but not all due to the limited experimental observations) of the mathematical model through calibration. Finally, we study the sensitivity of the parameters in our model.

The paper is structured as follows. The pore-scale model is defined in Sec. 2, where we introduce the basic concepts, ideas, and equations for modeling biofilms in the pore-scale. In Sec. 3 we describe the computational algorithm to solve numerically the model. In Sec. 4 we present different plots for some of the unknown model variables using the best available estimates for the input parameters. We perform a sensitivity analysis in Sec. 5 in order to detect the critical model parameters. Finally, in Sec. 6 we present the conclusions. 

\section{Pore-scale model}
In \cite{Noorden:Article:2010}, a pore-scale model for biofilm formation considering the biofilm as impermeable and formed by a single species is built. In \cite{Alpkvist:Article:2007}, a model for heterogeneous biofilm development considering the biofilm formed by different components is built. In this work we extend these ideas to build a model for biofilm formation which includes the notions of porosity and permeability.\par
We assume the following:
\begin{enumerate}
\item The biofilm is a separate phase (as being a porous medium itself), which is modeled by mass conservation and a growth potential.\label{A2}
\item The biofilm is modeled as a continuous medium consisting of four components: water, EPS, active bacteria, and dead bacteria.\label{A3}
\item The fluid flow and nutrients are in a steady state when we compute the biofilm growth potential and volumetric fractions at each time step.\label{A4}
\item The biofilm growth occurs in the lower substratum.\label{A5} 
\item There is only one nutrient, which is mobile both in the water and biofilm.\label{A1}
\item Temperature is constant (room temperature).\label{A7}
\item The gravity effects are neglected.\label{A8}
\item The bacterial growth rate is of Monod-type and the endogenous respiration is linear.\label{A9}
\end{enumerate}
We comment on the assumptions. Following \cite{Alpkvist:Article:2007}, we need \ref{A2} in order to properly model the dynamics of the biofilm components. The motivation for \ref{A4} is that the fluid mean flow and biofilm growth velocities are of orders mm/s and $10^{-5}$ mm/s, respectively (\cite{Duddu:Article:2009}). Then, water and nutrient displacements due to biofilm growth are neglected (\cite{Noorden:Article:2010}). We consider \ref{A5} because a T-microchannel is used to grow the biofilm, where bacteria and nutrients are first injected in the vertical channel and, afterwards, only nutrients are injected through the horizontal channel, leading to a greater growth of bacteria on the lower substrate (where the horizontal and vertical channels connect). \ref{A1} is taken for simplicity, but the extension to other nutrients can be achieved straightforwardly from the model equations. We consider \ref{A7} because the experiments are performed at room temperature. We consider \ref{A8} because in the experimental setting, the gravity direction is perpendicular to the plane where the biofilm grows. \ref{A9} is an experimental based way to model bacterial growth and death, which is commonly used in literature. We remark that unlike in simple biofilm models, we do not assume a constant density of the biofilm.
\subsection{Geometrical settings} 
\begin{figure}
\centering
\includegraphics{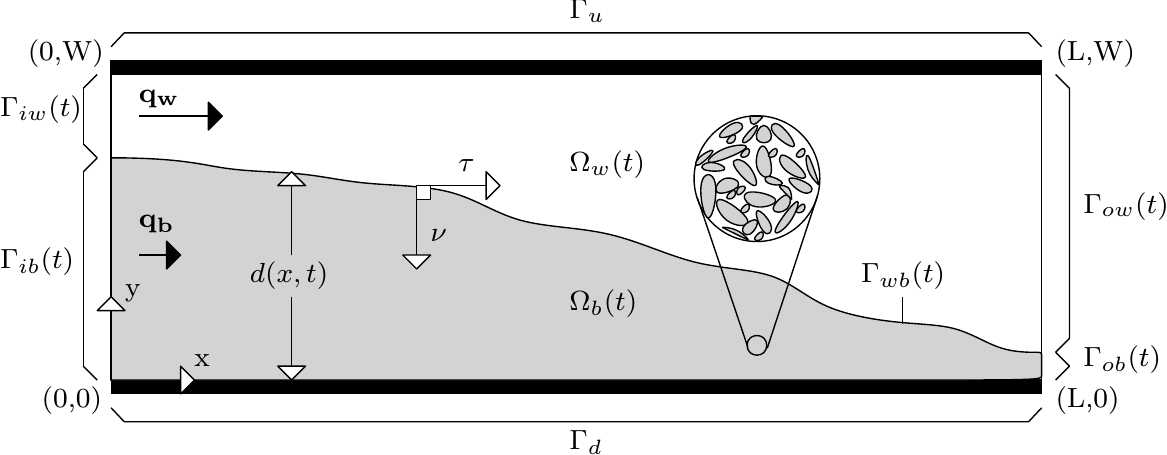}
\captionof{figure}{Schematic representation of the porous medium.}
\label{domains} 
\end{figure}

We consider a two-dimensional pore of length $L$ and width $W$:
\begin{equation*}
\Omega:=(0,L)\times (0,W).
\end{equation*}
The motivation for choosing this geometry is that we can approximate a porous medium in the macro scale as a bundle of tubes (\cite{Noorden:Article:2010}). Fig. \ref{domains} shows the water and biofilm domains and boundaries in the pore.\par
The boundary of the pore consists of the substrate, the inflow, and the outflow:
\begin{equation*}
\Gamma_u:=[0,L]\times \lbrace W\rbrace,\quad\Gamma_d:=[0,L]\times \lbrace 0\rbrace,\quad\Gamma_i:=\lbrace 0 \rbrace\times [0,W],\quad\Gamma_o:=\lbrace L \rbrace\times [0,W].
\end{equation*}
\\[5pt]
The domain of the pore consists of the biofilm and the water phase
\begin{align*}
&\Omega_b(t):=\lbrace (x,y)|0<x<L,\; 0<y<d(x,t)\rbrace,\\
&\Omega_w(t):=\lbrace (x,y)|0<x<L,\; d(x,t)<y<W \rbrace,
\end{align*}
where $d(x,t)$ is the biofilm thicknesses.\par
The interface between the water and biofilm phases is denoted by $\Gamma_{wb}(t)$, which mathematically is given by
\begin{align*}
&\Gamma_{wb}(t):=\lbrace (x,y)|0<x<L,\;y=d(x,t) \rbrace.
\end{align*}
The inflow and outflow boundaries for the water domain $\Omega_w(t)$ are given by
\begin{align*}
&\Gamma_{iw}(t):=\lbrace (x,y)|x=0,\;d(0,t)<y<W \rbrace,\\
&\Gamma_{ow}(t):=\lbrace (x,y)|x=L,\;d(L,t)<y<W\rbrace,
\end{align*}
while the inflow and outflow boundaries for the biofilm domain $\Omega_b(t)$ are given by
\begin{align*}
&\Gamma_{ib}(t):=\lbrace (x,y)|x=0,\;0<y<d(0,t) \rbrace,\\
&\Gamma_{ob}(t):=\lbrace (x,y)|x=L,\;0<y<d(L,t) \rbrace.
\end{align*}
The unit normal pointing into the biofilm and the tangential vector are given by
\begin{align*}
\pmb{\nu}=(\partial_x d,-1)^T/\sqrt{1+(\partial_x d)^2},\quad\pmb{\tau}=(1,\partial_x d)^T/\sqrt{1+(\partial_x d)^2}.
\end{align*}
In the next section, we define the equations for the flow, nutrients, and biofilm growth.
\subsection{Equations in the water phase}
The water is assumed to be incompressible. The water flow is described by the Stokes system
\begin{align*}
\nabla\cdot \textbf{q}_w=0,\quad\mu\Delta \textbf{q}_w=\nabla p_w,
 \end{align*}
where $\mu$ is the viscosity, $p_w$ is the water pressure, and $\textbf{q}_w=(q_w^{(1)},q_w^{(2)})$ is the water velocity.\par 

In the water phase, the nutrient concentration ($c_w$) satisfies the convection-diffusion equation
\begin{equation*}
\partial_t c_{w}+\nabla\cdot\pmb{J}_w=0,\quad \pmb{J}_w=-D\nabla c_{w}+\textbf{q}_wc_{w},
\end{equation*}
where $D$ and $\pmb{J}_w$ are the nutrient diffusion coefficient and nutrient flux in water, respectively.
\subsection{Equations in the biofilm phase}
As mentioned before, the biofilm components are: water, EPS, active bacteria, and dead bacteria ($j=\lbrace w,e,a,d\rbrace$). Let $\theta_j(t,\textbf{x})$ and $\rho_j(t,\textbf{x})$ denote the volume fraction and the density (relative to volume fraction) of species j at time $t$ and position $\textbf{x}$, respectively. The biomass and water are assumed incompressible ($\rho_j(t,\textbf{x})=\rho_j$). Therefore, the biofilm density in a given position and time is
\begin{equation*}
\rho(t,\textbf{x})=\sum_j\rho_j\theta_j(t,\textbf{x}).
\end{equation*} 
The volume fractions are constrained to
\begin{equation}
\sum_j\theta_j(t,\textbf{x})=1.
\label{constrain}
\end{equation}
In the biofilm, the biomass can increase or decrease due to EPS production, bacterial reproduction, and death of the bacteria. Let $\textbf{u}$ be the velocity of the biomass. Assuming that the biofilm growth is irrotational (\cite{Duddu:Article:2009}), we can derive the velocity field from a function potential $\Phi$:
\begin{equation*}
\textbf{u}=-\nabla\Phi.
\end{equation*}
In \cite{Hornung:Book:1997} the Brinkman model is derived as the Darcy scale counterpart of the Stokes model at the scale of pores, assuming that the volume of the porous media skeleton is much smaller than the volume of the reference cell. Therefore, recalling that biofilms are mostly water, we assume that the water content is constant ($\partial_t\theta_w=0$) and we describe the water flux in the biofilm by the mass conservation and the Brinkman equation
\begin{align*}
\nabla\cdot\textbf{q}_{b}=0,\quad\frac{\mu}{\theta_w}\nabla\vec{q}_b-\frac{\mu}{k}\vec{q}_{b}=\nabla p_{b},
\end{align*}
where $\textbf{q}_{b}$ and $p_{b}$ are the velocity and pressure of the water in the biofilm, respectively, and $k$ is the permeability. 

The conservation of mass for the biofilm components ($l=\lbrace e,a,d\rbrace$) is given by
\begin{equation}\label{compo}
\partial_t(\rho_l\theta_l)+\nabla\cdot(\textbf{u}\rho_l\theta_l)=R_l
\end{equation}
where $R_l$ are the rates on the volume fractions; these rates are discussed in more detail below.

Inside the biofilm, the nutrients are dissolved in the water. The nutrient concentrations satisfy the following convection-diffusion-reaction equations:
\begin{align*}
\partial_t(\theta_w c_{b})+\nabla\cdot\pmb{J}_b=R_{b},\quad \pmb{J}_b=-\theta_w D\nabla c_{b}+\textbf{q}_{b}c_{b},
\end{align*}
where $c_b$, $R_b$, and $\pmb{J}_b$ are the nutrient concentration, reaction term, and flux in the biofilm.

Following \cite{Alpkvist:Article:2007}, summing Eq. \ref{compo} over $l$ and using Eq. \ref{constrain} and $\rho_l$ are constants for all $l$, we obtain an expression for the growth velocity potential
\begin{equation*}
-\nabla^2\Phi=(1-\theta_w)^{-1}\sum_l\frac{R_l}{\rho_l}.
\end{equation*}
\subsection{Equations at the biofilm-water interface}
Coupling conditions for free flow and flow in a porous media is an active research topic and there are several works that study this problem (\cite{Beavers:Article:1967,Saffman:Article:1971,Urquiza:Article:2008,Dumitrache:Article:2012,Yang:Article:2017}). We assume that the normal velocity of the interface between the biofilm and fluid is negligible with respect to the velocity of the fluid phase (\cite{Noorden:Article:2010}). Then, we choose conditions of continuous velocity and continuity of the normal component of the stress tensor (\cite{Dumitrache:Article:2012})
\begin{align*}
\vec{q}_{w}=\vec{q}_{b},\quad\pmb{\nu}\cdot (\mu\nabla\vec{q}_{w}-\mathbb{1}p_w)=\pmb{\nu}\cdot ((\mu/\theta_w)\nabla\vec{q}_{b}-\mathbb{1}p_b).
\end{align*}
Conservation of nutrients is ensured by the Rankine-Hugoniot condition:
\begin{align*}
(\pmb{J}_b-\pmb{J}_w)\cdot \pmb{\nu}=\nu_n(\theta_wc_b-c_w).
\end{align*}
The nutrient concentration is assumed continuous across the interface:
\begin{align*}
\theta_wc_{b}=c_{w}.
\end{align*}
We set the growth velocity potential at the interface to zero:
\begin{equation*}
\Phi=0.
\end{equation*}
The location of the interface $\Gamma_{wb}(t)$ changes in time due to the production of EPS, active bacteria, death of the active bacteria, and shear stress produced by the water flux. In \cite{Horn:Article:2014}, the authors write a review of modeling of biofilm systems, which includes a summary of different detachment models. However, none of those detachment models are given as a function of the flow velocity. To incorporate this, we follow  \cite{Taylor:Article:1990} and \cite{Noorden:Article:2010} and use the following definition for the tangential shear stress:
\begin{equation*}
S=||(\mathbb{1}-\pmb{\nu}{\pmb{\nu}}^T)\mu(\nabla \textbf{q}_w+\nabla \textbf{q}_w^T)\pmb{\nu}||,
\end{equation*}
where the norm that we use is the maximum norm. Then, the normal velocity of the interface is given by
\begin{equation*}
\nu_n=\begin{cases}
[\pmb{\nu}\cdot\vec{u}]_{+},& d=W,\\
\pmb{\nu}\cdot\vec{u}+k_{str}S,&0<d<W,\\
0,&d=0,\end{cases}
\end{equation*}
where $k_{srt}$ is a constant for the shear stress. In the above, we ensure that the interface does not cross the strip by taking the positive cut on the right-hand side when $d=W$, which means that only death of active bacteria would lead to the biofilm thickness to decrease. Following \cite{Noorden:Article:2010}, the evolution equation for the biofilm thickness reads as
\begin{equation*}
\partial_t d=\begin{cases}
-\sqrt{1+(\partial_x d)^2}[\pmb{\nu}\cdot\vec{u}]_{+},& d=W,\\
-\sqrt{1+(\partial_x d)^2}(\pmb{\nu}\cdot\pmb{u}+k_{str}S),&0<d<W,\\
0,&d=0.\end{cases}
\end{equation*}
Finally, homogeneous Neumann condition is considered for the biofilm components: 
\begin{equation*}
\pmb{\nu}\cdot\nabla\theta_l=0.
\end{equation*}
\subsection{Boundary and initial conditions}
At the inflow, we specify the pressure and nutrient concentration and we consider homogeneous Neumann condition for the growth velocity potential and volumetric fractions:
\begin{alignat*}{2}
p_w=p_i,\quad c_w=c_i\quad\text{at}\;\Gamma_{iw},\\
p_b=p_i,\quad c_b=c_i/\theta_w,\quad \pmb{\nu}\cdot \nabla\Phi= \pmb{\nu}\cdot \nabla\theta_l=0\quad\text{at}\;\Gamma_{ib}.
\end{alignat*}
At the outflow, we specify the pressure and we consider Neumann conditions for the concentrations, growth velocity potential, and volumetric fractions:
\begin{alignat*}{2}
p_w=p_o,\quad\pmb{\nu}\cdot \nabla c_w=0&\quad\text{at}\;\Gamma_{ow},\\
p_b=p_o,\quad\pmb{\nu}\cdot \nabla c_b=\pmb{\nu}\cdot \nabla\Phi=\pmb{\nu}\cdot \nabla\theta_l=0&\quad\text{at}\;\Gamma_{ob}.
\end{alignat*}
At the lower substrate, we consider a no-flux boundary condition for the water, nutrients and volumetric fractions, and homogeneous Neumann condition for the growth potential:
\begin{align*}
\pmb{\nu}\cdot \textbf{q}_b=\pmb{\nu}\cdot \textbf{J}_b=\pmb{\nu}\cdot \nabla\theta_l= \pmb{\nu}\cdot \nabla\Phi=0\quad\text{at}\;\Gamma_{d}.
\end{align*}
At the upper substrate, we consider a no-slip boundary condition for the free flow and no-flux for the nutrient concentration:
\begin{align*}
q_w^{(1)}=q_w^{(2)}=\pmb{\nu}\cdot \textbf{J}_w=0\quad\text{at}\;\Gamma_{u}.
\end{align*}
The initial pressure, nutrient concentrations, growth potential, biofilm height, and volume fractions are given.
\subsection{Reaction terms}
The bacteria needs to consume nutrients in order to produce EPS and for reproduction. We model this using Monod-type functions \cite{Horn:Article:2014}. Also, we consider a linear death rate of bacteria. Then, we have the following reaction terms:
\begin{align*}
R_b&=-\mu_n\theta_a\rho_a\frac{c_{b}}{k_n+c_{b}},\\
R_e&=Y_e\mu_n\theta_a\rho_a\frac{c_{b}}{k_n+c_{b}},\\
R_a&=Y_a\mu_n\theta_a\rho_a\frac{c_{b}}{k_n+c_{b}}-k_{\text{res}}\theta_a\rho_a,\\
R_d&=k_{\text{res}}\theta_a\rho_a,
\end{align*}
where $Y_e$ and $Y_a$ are yield coefficients, $\mu_n$ is the maximum rate of nutrient utilization, $k_n$ is the Monod half nutrient velocity coefficient, and $k_{\text{res}}$ is the endogenous respiration rate. 
\subsection{Pore-scale model for permeable biofilm}
For increasing the readability of the paper we summarize here the developed mathematical model for permeable biofilm:
\begin{center}
\begin{tabular}{l l l}
&\textbf{Water flow:}&\\
Stokes equations&$\nabla\cdot \textbf{q}_w=0\quad\mu\Delta \textbf{q}_w=\nabla p_w$&$\Omega_w(t)$.\\
Continuity velocities&$\textbf{q}_{w}=\textbf{q}_{b}$& $\Gamma_{wb}(t)$.\\
Continuity stress tensor&$\pmb{\nu}\cdot (\mu\nabla\vec{q}_{w}-\mathbb{1}p_w)=\pmb{\nu}\cdot ((\mu/\theta_w)\nabla\vec{q}_{b}-\mathbb{1}p_b)$& $\Gamma_{wb}(t)$.\\
Brinkman equations&$\nabla\cdot\textbf{q}_{b}=0\quad(\mu/\theta_w)\nabla\vec{q}_b-(\mu/k)\vec{q}_{b}=\nabla p_{b}$ & $\Omega_{b}(t)$.\\ 
&\textbf{Nutrient transport:}&\\
Conservation of mass&$\partial_t c_{w}+\nabla\cdot\pmb{J}_w=0$ & $\Omega_w(t)$.\\
Rankine-Hugoniot&$(\pmb{J}_b-\pmb{J}_w)\cdot \pmb{\nu}=\nu_n(\theta_wc_b-c_w)$& $\Gamma_{wb}(t)$.\\
Continuity of nutrients&$\theta_wc_b=c_w$& $\Gamma_{wb}(t)$.\\
Conservation of mass&$\partial_t(\theta_w c_{b})+\nabla\cdot\pmb{J}_b=R_{b}$ & $\Omega_{b}(t)$.\\ 
&\textbf{Growth velocity potential:}&\\
Reference potential&$\Phi=0$& $\Gamma_{wb}(t)$.\\
Potential equation&$-\nabla^2\Phi=\Sigma_l(R_l/\rho_l)/(1-\theta_w)\quad \pmb{u}=-\nabla\Phi$&$\Omega_{b}(t)$.\\ 
&\textbf{Volume fractions:}&\\
Detached component&$\pmb{\nu}\cdot\nabla\theta_l=0$ & $\Gamma_{wb}(t)$.\\
Conservation of mass&$\partial_t\theta_l+\nabla \cdot (\pmb{u}\theta_l)=R_l/\rho_l$ & $\Omega_{b}(t)$.\\ 
&\textbf{Biofilm-water interface:}&\\ 
Biofilm thickness &$\partial_t d=\begin{cases}
-\sqrt{1+(\partial_x d)^2}[\pmb{\nu}\cdot\vec{u}]_{+},& d=W,\\
-\sqrt{1+(\partial_x d)^2}(\pmb{\nu}\cdot\pmb{u}+k_{str}S),&0<d<W,\\
0,&d=0,\end{cases}$&$\Gamma_{wb}(t)$.\\ 
&\textbf{Reaction terms:}&\\
Nutrient consumption&$R_b=-\mu_n\theta_a\rho_ac_{b}/(k_n+c_{b})$&$\Omega_{b}(t)$.\\
Death of bacteria&$R_d=k_{\text{res}}\theta_a\rho_a$&$\Omega_{b}(t)$.\\
Bacterial reproduction&$R_a=-Y_aR_b-R_d$&$\Omega_{b}(t)$.\\
EPS production&$R_e=-Y_eR_b$&$\Omega_{b}(t)$.
\end{tabular}
\end{center}

\noindent The aforementioned equations define the pore-scale model for permeable biofilm.  This is a coupled system of nonlinear partial differential equations with a moving interface.

We use an ALE method for tracking the biofilm-water interface (\cite{Donea:Article:2004}). We use backward Euler for the time discretization and linear Garlekin finite elements for the spatial discretization. We split the solution process into three sub-steps. A damped version of Newton's method is used in each of the steps. First, we solve for the pressures and water fluxes. Secondly, we solve for the nutrient concentration. Thereafter, we solve for the volumetric fractions, growth potential, and biofilm thickness. We iterate between the previous steps until the error (the difference between successive values of the solution) drops below a given tolerance. Then, we move to the next time step and solve again until a given final time. We implement the model equations in the commercial software COMSOL Multiphysics (COMSOL 5.2a, Comsol Inc, Burlington, MA, www.comsol.com).

\section{Model test}\label{Model test}
\begin{figure}
\centering
\includegraphics[scale=.7]{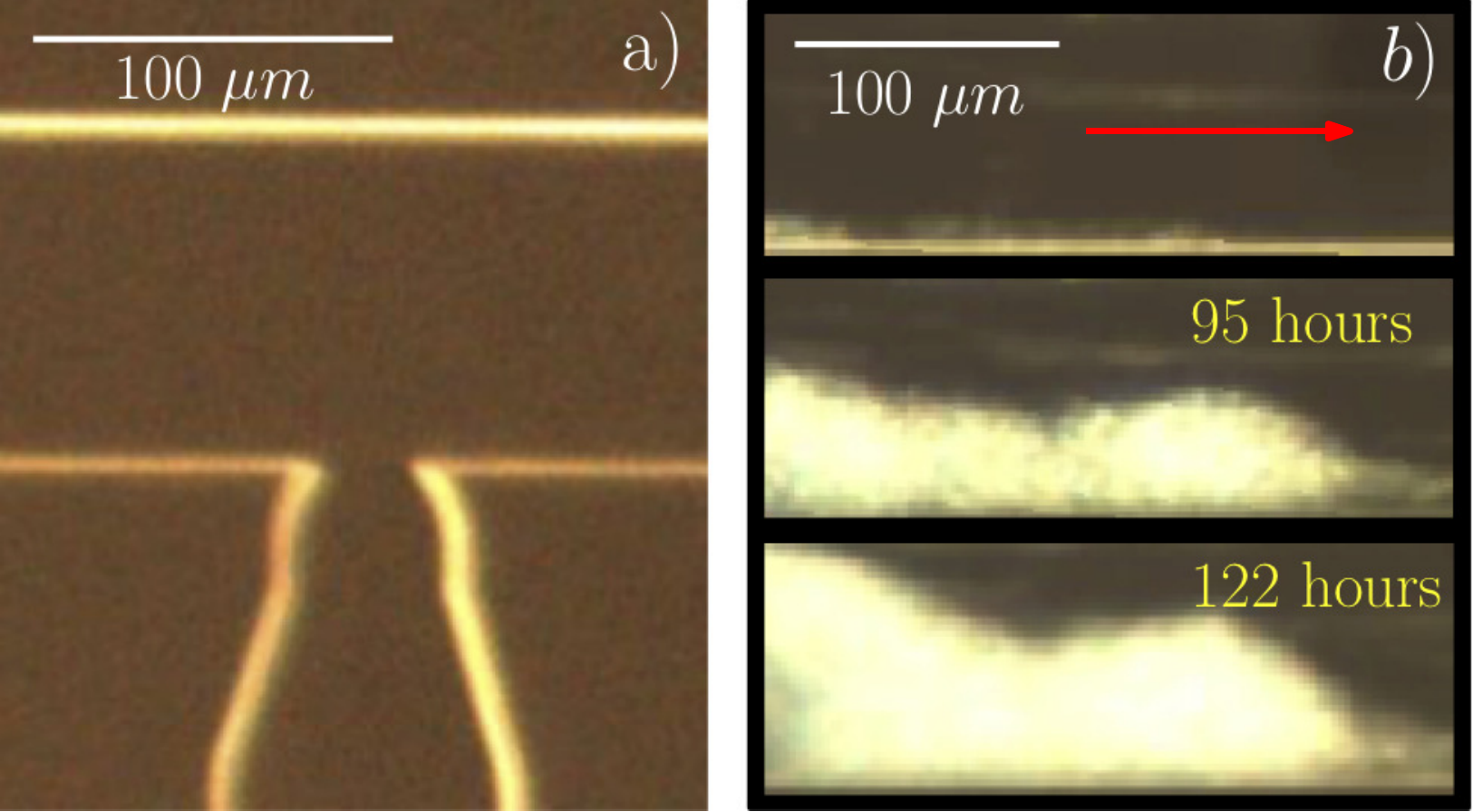}
\captionof{figure}{a) T-microchannel and b) biofilm formation.}
\label{hor} 
\end{figure}

Micro model experiments under controlled conditions have been designed for allowing for determination of critical input parameters for the biofilm formation. A glass micromodel (Micronit, Netherland)
a camera (VisiCam 5.0), and two syringe pumps (NE-1000 Series, Syringe Pumps) were used to perform the experiments. We managed to establish biofilm growth in the glass micromodel and studied the different biofilm growth profiles varying the water flux. The micromodel used in the laboratory has a width of 100 $\mu$m and thickness/depth of 20 $\mu$m. Fig. \ref{hor} shows the biofilm formation over time for a flow rate of $0.2\;\mu l/min$, which corresponds to a water velocity injection of $q_i=1.66 \;\text{mm/s}:=U$ and an entry pressure of $p_{i}=0.128\;Pa$. First, microbes and nutrients were injected in the vertical channel for 24 hours at a rate of 1 $\mu l/\text{min}$. Afterwards, the vertical channel was closed for one day. Then, we started to inject nutrients from the left channel at a rate of 0.2 $\mu l/\text{min}$. The injected nutrient concentration was $c_i=0.88\;kg/m^3$. A detailed description of the performed experiments can be found in \cite{Liu:Article:2018}.\\[10pt] 
\noindent
\\[10pt]
In order to compare the mathematical model with the laboratory experiments, we perform numerical simulations considering the same experimental input values for flux and nutrient concentration. We study the increase of percentage of biofilm coverage area over time. We consider a space domain of the same width of the micro channel $W=0.1$ mm and length $L=0.2$ mm. Recalling that biofilms are mostly composed by water, we set the water volume fraction in the biofilm equal to 90\% ($\theta_w=0.9$). Then, the organic matter in the biofilm is equal to 10\%. We assume that initially the biomass in the biofilm is formed only by active bacteria ($\theta_a(0,x,y)=0.1,\;\theta_e(0,x,y)=0,\;\text{and}\;\theta_d(0,x,y)=0$). We set the initial biofilm thickness to $d(0,x)=2.5\; \mu m$. A combination between experimentally determinated parameters and values from literature has been used for the numerical simulations, see Table \ref{parameters} for details.
\begin{table}[h]
\centering
\label{tab:1}
\caption{Table of model parameters for the verification study}
\begin{tabular}{ l l l l}		
\hline
Name & Description & Value & Refs.\\
\hline
$k_\text{res}$ & Bacterial decay rate&	$2\times 10^{-6}/\text{s}$&\cite{Alpkvist:Article:2007}\\ 
$\mu_n$ & Maximum growth rate &  $10^{-5}/\text{s}$&\cite{Alpkvist:Article:2007}\\
$k_n$	& Monod-half velocity&	$10^{-4}\;\text{kg}/\text{m}^{3}$&\cite{Alpkvist:Article:2007}\\
$D$	& Nutrient diffusion coefficient&	$1.7\times 10^{-9}\;\text{m}^2/\text{s}$ &\cite{Duddu:Article:2009}\\
$\rho_e$	& EPS density &	$1012.5\;\text{kg}/\text{m}^{3}$	 & \cite{Duddu:Article:2009}\\
$\rho_a$	& Active bacterial density &	$1025\;\text{kg}/\text{m}^{3}$	 & \cite{Duddu:Article:2009}\\
$\rho_d$	& Dead bacterial density &	$1025\;\text{kg}/\text{m}^{3}$	 & \cite{Duddu:Article:2009}\\
$Y_{a}$ & Active bacterial growth yield	&	$.553$&\cite{Duddu:Article:2009}\\
$Y_{e}$	& EPS growth yield&	$.447$&\cite{Duddu:Article:2009}\\
$\mu$	& Water dynamic viscosity &	$10^{-3}\;\text{Pa}\cdot\text{s}$ & \cite{Crittenden:Book:2012}\\
$\rho_w$	& Water density &	$10^3\;\text{kg}/\text{m}^{3}$&\cite{Crittenden:Book:2012}\\
$k$&Biofilm permeability&$10^{-10}\; m^2$& \cite{Deng:Article:2013}\\
\hline
\end{tabular}
\label{parameters}
\end{table}

\begin{figure}
\centering
\includegraphics[scale=.48]{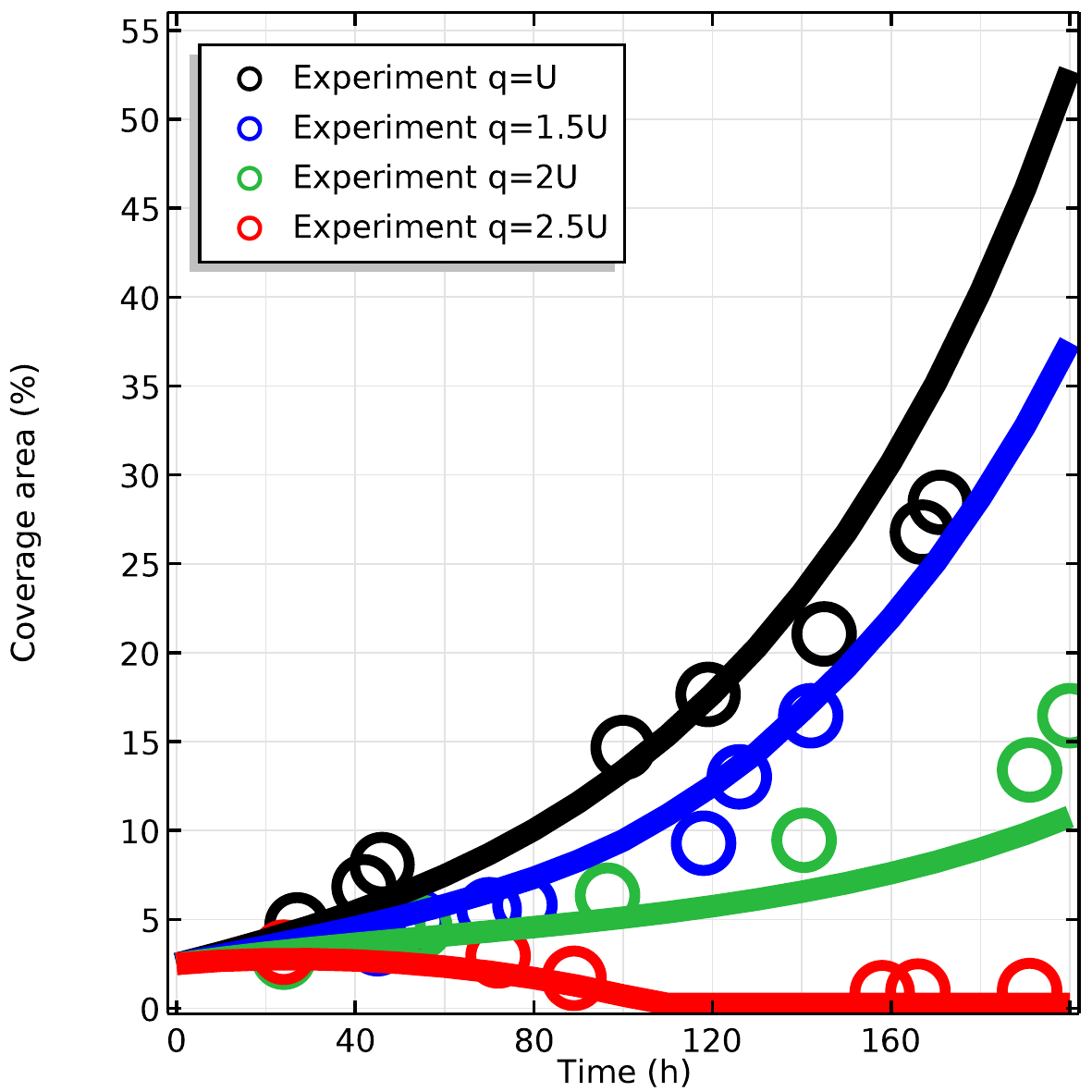}
\captionof{figure}{Experimental data and numerical simulations for 4 different flux conditions.}
\label{thi} 
\end{figure}

For calibration of the stress coefficient, we consider the experimental percentage of biofilm area over time for four different water velocities. Fig. \ref{thi} shows the experimental and simulated percentage of biofilm area over time. After numerical simulations, the order of stress coefficient that best fits the data is $k_{str}=10^{-10}\;\text{m}/(\text{s Pa})$. Then, we perform a parametric sweep of the stress coefficient in the interval $[10^{-9},10^{-11}]$ with a step of $10^{-11}$, where we use the method of least squares. The value that best fits the experimental data is $k_{str}=2.6\times 10^{-10}\;\text{m}/(\text{s Pa})$.\\

\noindent
\\[5 pt]
\section{Numerical results}\label{Numerical results}
We perform numerical simulations with $c_i=1\times 10^{-3}\;kg/m^3$, $p_{i}=0.128\;Pa$, $\theta_w=0.9$, and $d(0,x)=2.5\; \mu m$. We consider a smaller nutrient concentration in comparison to the one used in the laboratory experiments to study the biofilm dynamics with nutrient limitation. We consider a heterogeneous biofilm, where initially the biomass on the left half side ($0<x<L/2$) is formed by 60$\%$ of active bacteria and 40$\%$ of EPS and the biomass on the right half side ($L/2<x<L$) is formed by 40$\%$ of active bacteria and 60$\%$ of EPS. The remaining input parameters are taken from Table \ref{parameters} and the calibrated stress coefficient is $k_{str}=2.6\times 10^{-10}\;\text{m}/(\text{s Pa})$. In the next figures, different numerical results at different times are shown. 

\begin{figure}
\centering
\includegraphics[scale=.48]{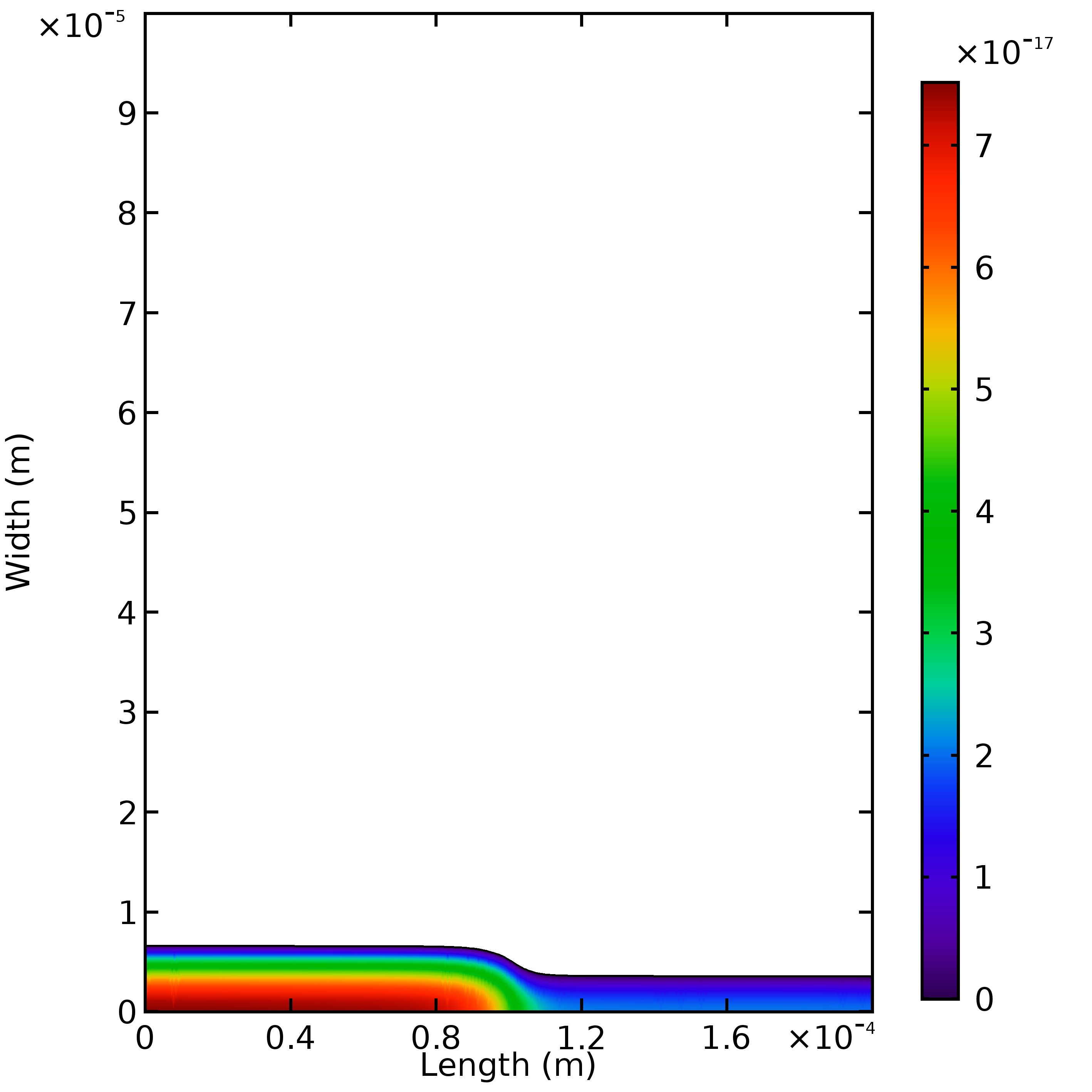}\includegraphics[scale=.48]{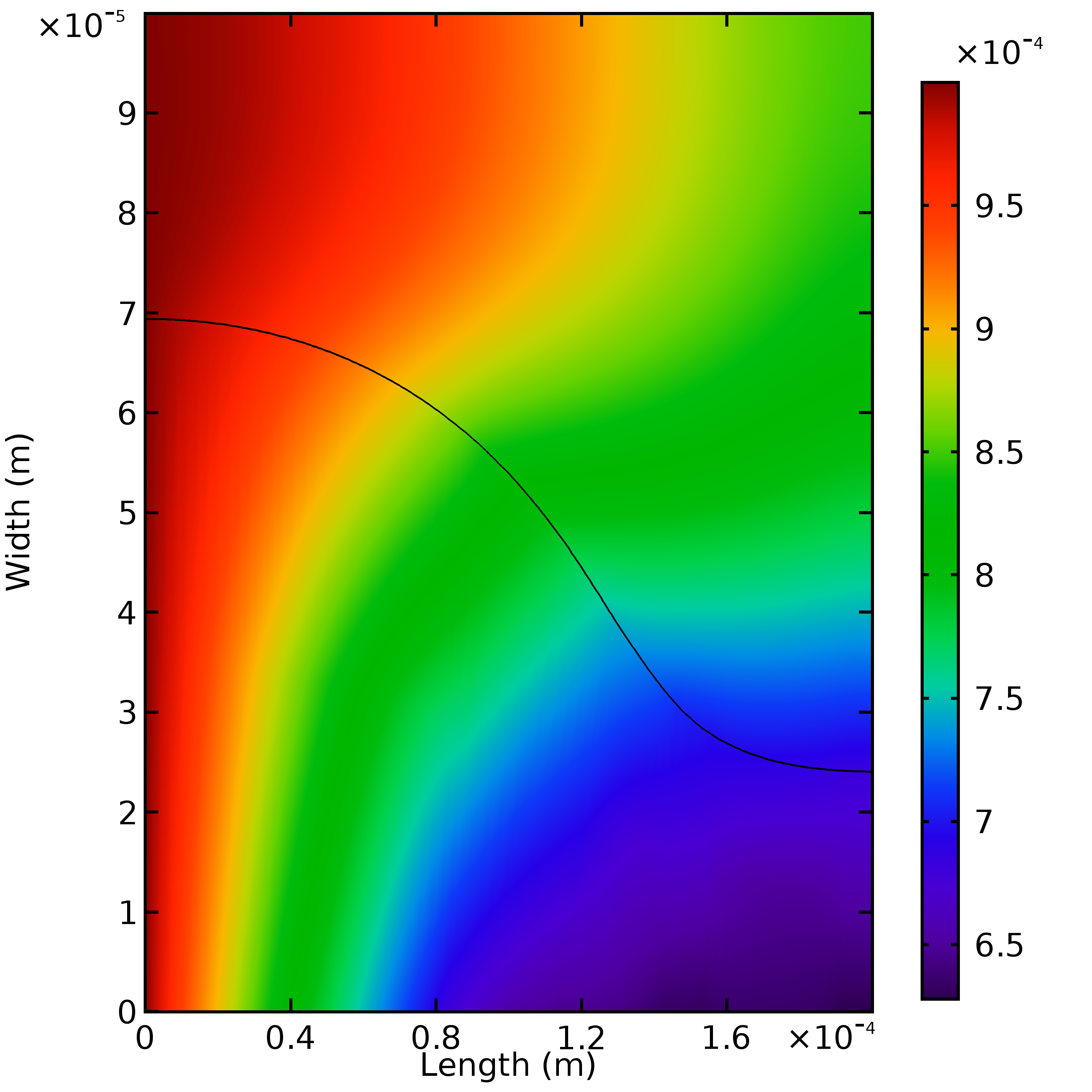}
\captionof{figure}{Growth velocity potential after 120 hours (left) and nutrient concentration after $360$ hours (right).}
\label{nutrient} 
\end{figure}

Fig. \ref{nutrient} shows the growth velocity potential $\phi$ after 120 hours and the nutrient concentrations $c_b$ and $c_w$ after 360 hours respectively. The growth velocity potential is larger on the left lower corner, as a result of the nutrient injection on the left side and the condition of zero potential on the interface. Therefore, the biomass will grow towards the right upper corner. After 360 hours of injection of nutrients, we observe that the nutrient concentration in the biofilm decreases from left to right, due to the consumption of nutrients by the active bacteria.\\[5 pt]
\noindent

\begin{figure}
\centering
\includegraphics[scale=.48]{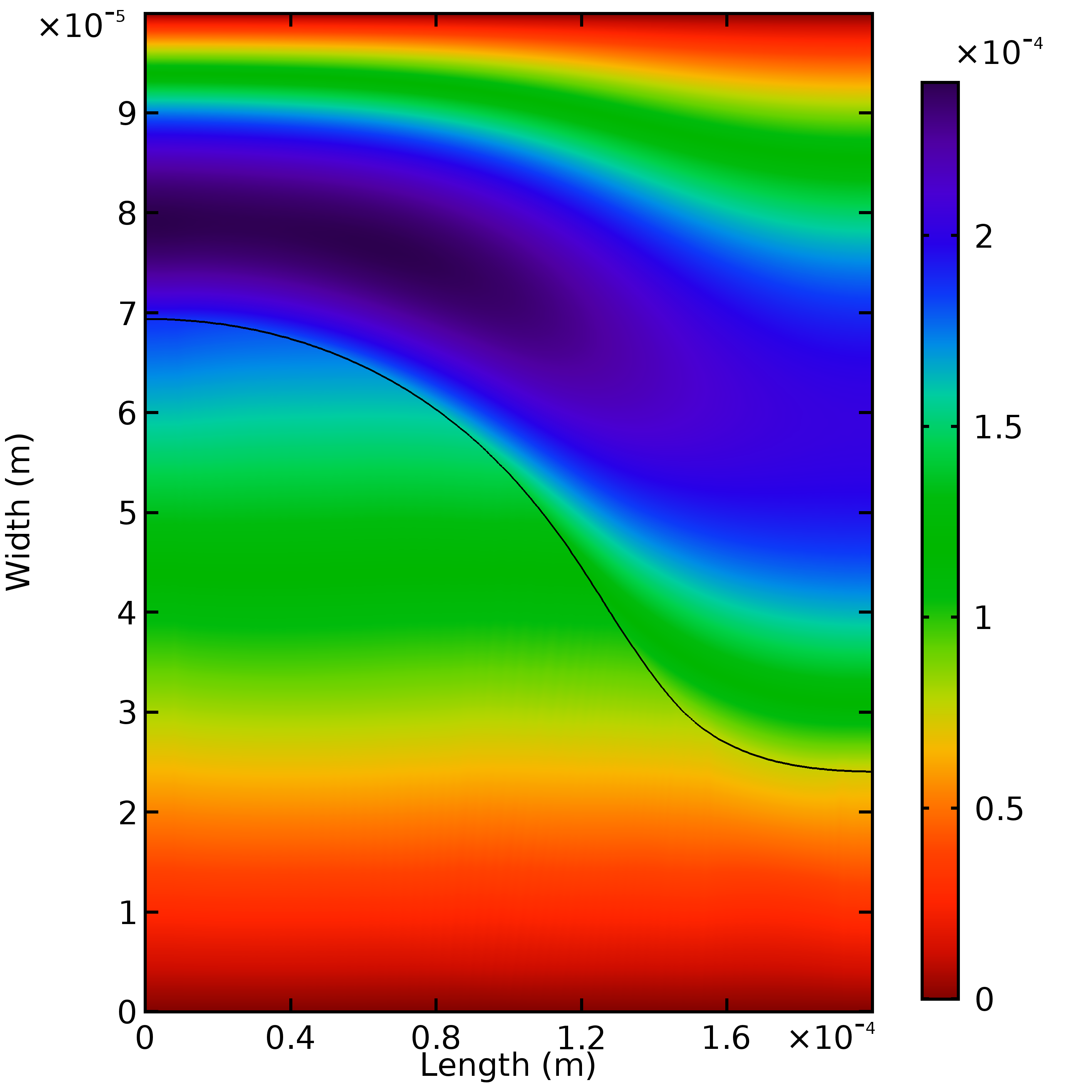}\includegraphics[scale=.48]{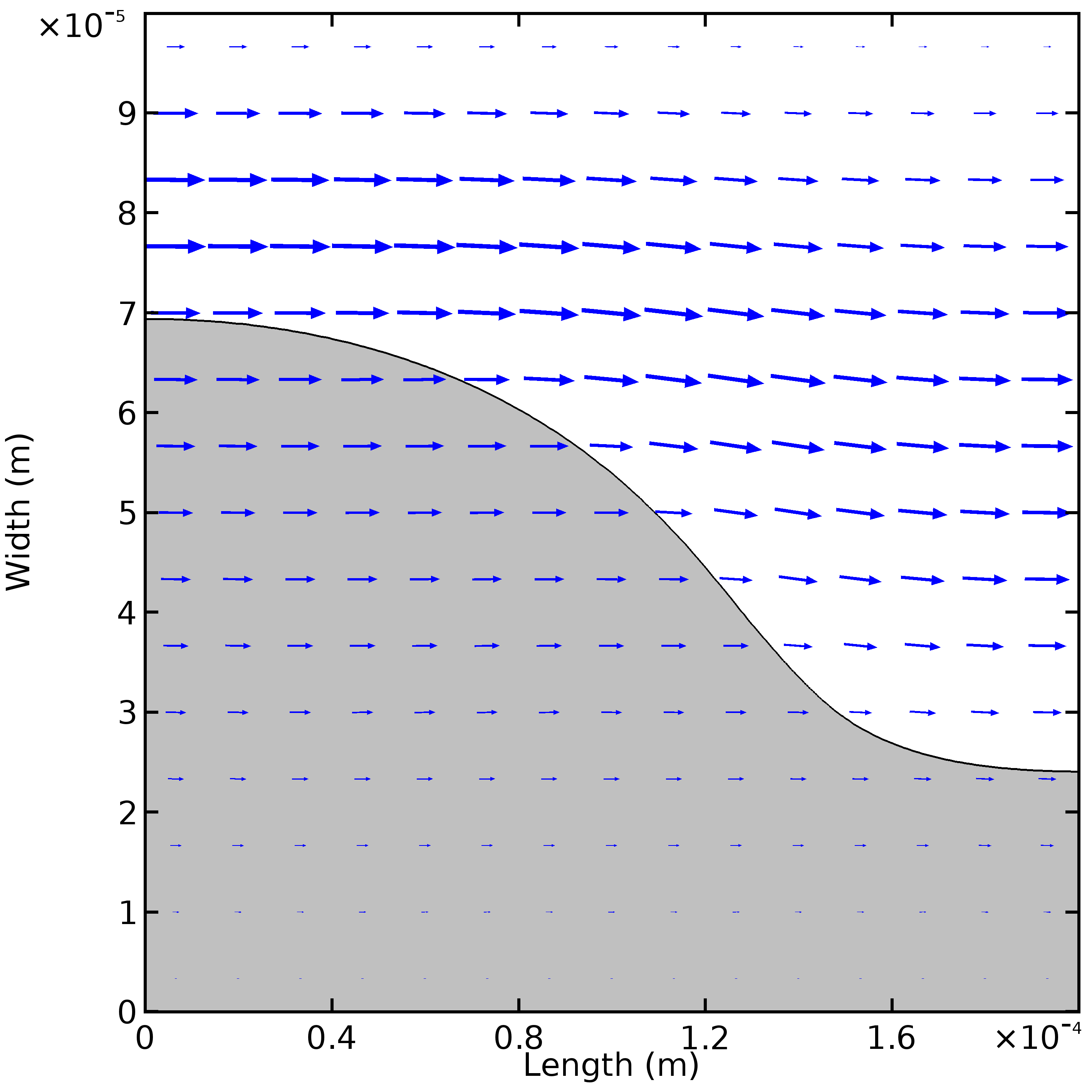}
\captionof{figure}{Magnitude (left) and direction (right) of the water flux velocity after 360 hours.}
\label{velocity} 
\end{figure}

Fig. \ref{velocity} shows the magnitude and the flow direction of the water flux velocity after $360$ hours respectively. In the water domain, we observe that the water flux is larger between the wall and the interface. In the biofilm, the water flux decreases from the interface until zero on the wall.

\begin{figure}
\centering
\includegraphics[scale=.48]{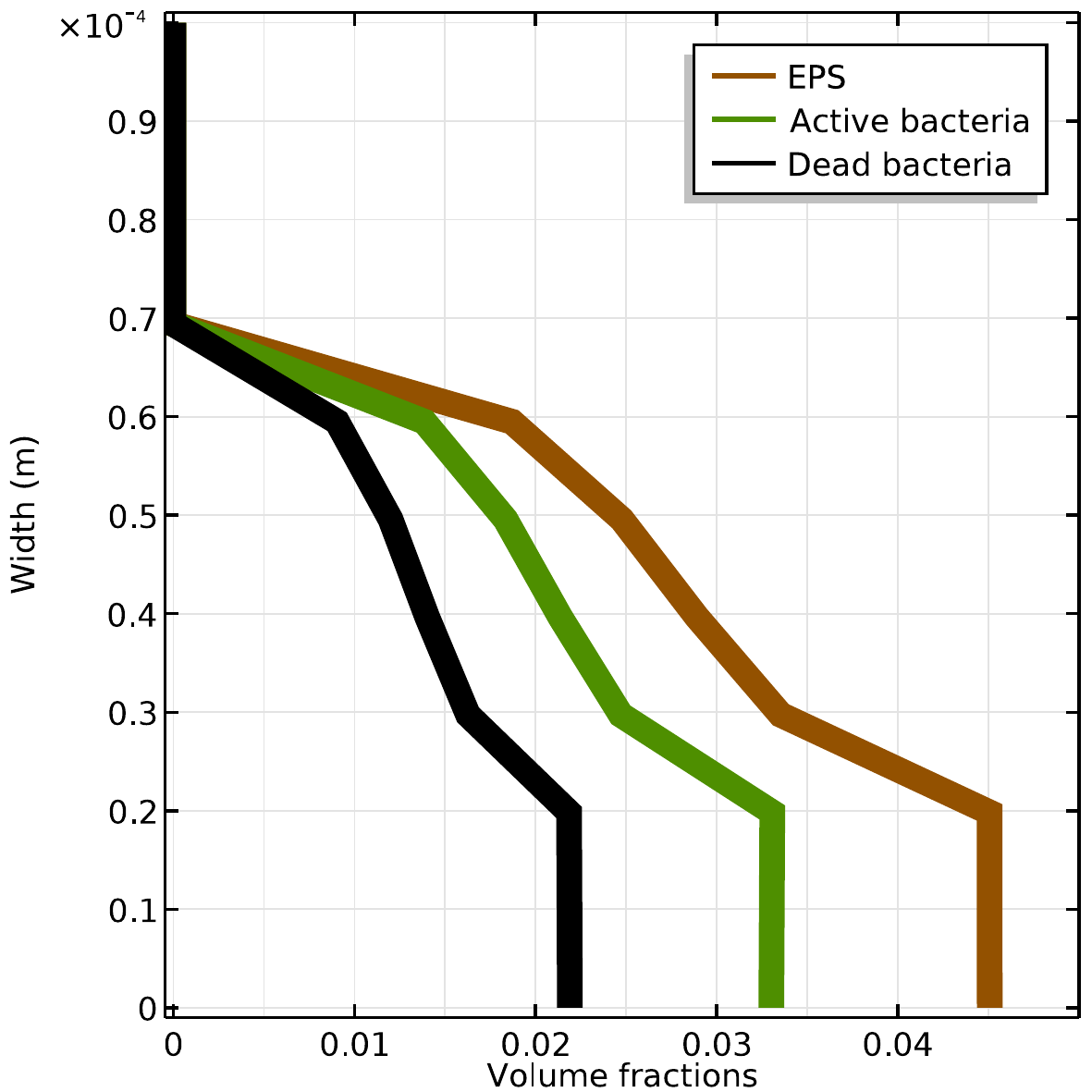}\includegraphics[scale=.48]{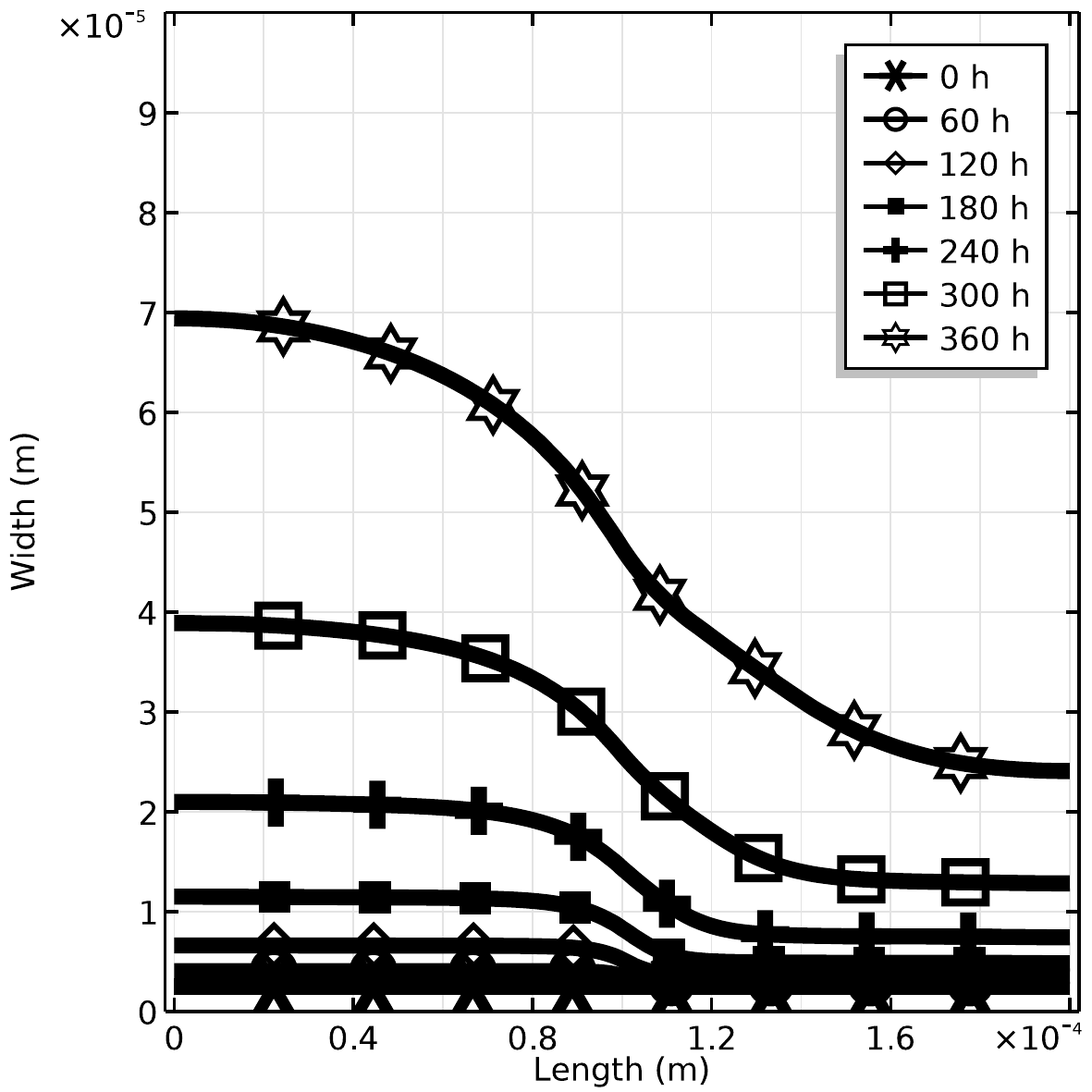}
\captionof{figure}{Total volumetric fractions after 360 hours (left) and biofilm height profile over time (right).}
\label{fracc} 
\end{figure}

Fig \ref{fracc} shows the total volumetric fraction after 360 hours and the biofilm height profile over time respectively. We observe that  more than 65\% of the organic matter in the biofilm is formed by EPS and dead bacteria after 360 hours. We also observe that the biofilm height on the left side grows faster over time, due to the nutrients being injected on the left side and also due to the larger initial active bacteria on the left half side, leading to a faster EPS and bacterial production.\\[5 pt]
\noindent
\\[5 pt]
\noindent
\\[5 pt]

\section{Sensitivity analysis}
Variability in input parameters may have a significant effect on output quantities of interest, for instance the percentage (0-100\%) of biofilm area relative to the area of the whole domain. We perform a global sensitivity analysis (\cite{Sobol:Article:2001,Sudret:Article:2008}) to quantify the effect of variability or uncertainty in ten material parameters that are assumed to be sensitive with respect to variation in the biofilm area after $T=50\;h$ of nutrient injection. The initial biofilm thickness is $d(0,x)=10\;\mu m$, the injected nutrient concentration is $c_i=0.88\;kg/m^3$, and the entry pressure is $p_i=0.128\;Pa$. The input parameters, their range of variation, and total Sobol index are listed in Table \ref{tab:sens_results} (see Appendix A for details).\par
\begin{table}[h!]
\centering
\caption{Total contribution of each material parameter on the relative variability of the biofilm area. Total effect sums to 1.38.}
\begin{tabular}{l  l  c c}
\hline
Parameter & Symbol & Range & Total Sobol Index\\
\hline
Diffusion coefficient & $D$ & $[1.53, \ 1.87]\times 10^{-9}$ & 0.126 \\
Monod-half velocity & $k_n$ & $[0.9, \ 1.1]\times10^{-4}$ & 0.0531 \\
Active bacteria yield & $Y_a$ & $[4.98, \ 6.08] \times 10^{-1}$ & 0.2188\\
Decay rate & $k_{res}$ & $[1.8, \ 2.2]\times10^{-6}$ &  0.0625 \\
Maximum growth rate & $\mu_{n}$ & $[0.9, \ 1.1]\times10^{-5}$ &  0.4139 \\
Stress & $k_{str}$ & $[2.34, \ 2.86]\times10^{-10}$ & 0.1582  \\
Permeability & $k$ & $[0.9, \ 1.1]\times10^{-10}$&  0.0675 \\
Water vol. fraction & $\theta_{w}$ & $[8.1, \ 9.9]\times10^{-1}$& 0.0952\\
Bacterial density & $\rho_{a}$ & $[0.9225, \ 1.1275]\times10^{3}$& 0.0924\\
EPS density & $\rho_{e}$ & $[0.91125, \ 1.11375]\times10^{3}$& 0.0926\\
\hline
\end{tabular}
\label{tab:sens_results}
\end{table}

Due to interaction between the parameters and the fact that the total contribution from a given parameter also involves all combinations of this parameter together with the other parameters, the sum of the relative total contribution from the parameters exceeds 1. The relative variability contribution from each parameter is significant for the parameter ranges investigated. The maximum growth rate stands out as more important than the others with respect to total variability, but none of them should be discarded based on this numerical sensitivity study alone. The true value of each of the ten parameters should be estimated with sufficient accuracy to lead to a reliable estimate of the biofilm area.

All previous plots are the result of the parameter values, initial conditions, and input values. The concept of growing a biofilm in the laboratory seems uncomplicated. Nevertheless, the biofilm formation takes up to two weeks and it is very sensitive to the surrounding conditions (e.g., the substrate surface and light conditions). As a result of limitations in the laboratory, we could not estimate all model parameters from the experiments. Then, it is necessary to improve the growth techniques and develop new measurement strategies to give better estimates of the parameters and, in turn, validate the model assumptions.
\section{Conclusions}
In this work, a pore-scale model for biofilm formation is built considering the biofilm as a porous medium. To our knowledge, the present work is the first study considering a permeable biofilm in a strip geometry. The stress coefficient $k_{str}=2.6\times 10^{-10}\;\text{m}/(\text{s Pa})$ is selected to match the experimental results. A sensitivity analysis is performed. The sensitivity analysis confirms that the variability or uncertainty in none of the 10 studied parameters should be neglected. In the numerical simulations, we observe a reduction of the biofilm height as the water flux velocity increases. For high flow rates we must consider the effects of the flow inside the biofilm, which affect the transport of nutrients and, therefore, influence the biofilm thickness.\\[10pt]

\noindent\textbf{Acknowledgements} The work of DLM, NL, KK, PP, GB, TS, and FAR was partially supported by GOE-IP and the Research Council of Norway through the projects IMMENS no. 255426 and CHI no. 255510. ISP was supported by the Research Foundation-Flanders (FWO) through the Odysseus programme (project G0G1316N) and by Statoil through the Akademia grant.
\section*{Appendix A: Sensitivity analysis method}
In this Appendix, we describe the theory behind the performed sensitivity analysis. The variation is assumed uniform in the sense that each parameter varies within a range where all values are equally likely. The sensitivity analysis relies on the Hoeffding or Sobol decomposition of the quantity of interest, here denoted $q$, as a series expansion in subsets of the $n$ input parameters $\pmb{y} = (y_1,...,y_n)$,
\begin{equation*}
q(\pmb{y}) = q^{\lbrace \emptyset \rbrace} + \sum_{i=1}^{n}q^{\lbrace i\rbrace}(y_i) + \sum_{i=1,j>i}^{n} q^{\lbrace i,j\rbrace}(y_i,y_j)+\ldots + q^{\lbrace 1,\ldots, n\rbrace}(\pmb{y}).
\end{equation*}
The Sobol decomposition terms are defined recursively as integrals over subsets of the range of $\pmb{y}$, denoted $\pmb{Y}$. We introduce a uniform weight function $w(\pmb{y})=w_1(y_1)...w_{n}(y_n)$ with $w_i = 1/(\max(y_i)-\min(y_i))$ and the subscript notation $\sim i$ to denote all parameters except parameter $i$. The decomposition terms are then determined by
\begin{align*}
q^{\lbrace \emptyset \rbrace}  &= \int\limits_{\pmb{Y}}q(\pmb{y})w(\pmb{y})d\pmb{y},\\
q^{\lbrace i\rbrace}(y_i)  &= \int\limits_{\pmb{Y}_{\sim i}}q(\pmb{y})w_{\sim i}(\pmb{y}_{\sim i})d\pmb{y}_{\sim i} - q^{\lbrace \emptyset \rbrace}, \quad 1\leq i \leq n,  \\
q^{\lbrace i, j\rbrace}(y_i,y_j)  &= \int\limits_{\pmb{Y}_{\sim i, j}}q(\pmb{y})w_{\sim i,j}(\pmb{y}_{\sim i,j})d\pmb{y}_{\sim i,j} - q^{\lbrace i\rbrace}(y_i)  - q^{\lbrace j \rbrace}(y_j) - q^{\lbrace \emptyset \rbrace},\;1\leq i < j \leq n 
\end{align*}
and so on for higher-order terms. 

The Sobol index for the $s$-parameter combination $\{ y_{i_1},y_{i_2},...,y_{i_s}\}$ is given by
\[
S_{ \{ i_1,...,i_s \rbrace} = \frac{1}{Var(q)}\int\limits_{\pmb{Y}_{i_1,...,i_s}} (q^{\lbrace i_1,...,i_s\rbrace}(y_{i_1},...,y_{i_s}))^2 w_{i_1}(y_{i_1})...w_{i_s}(y_{i_s}) dy_{i_1}\hdots dy_{i_s}.
\]
The total variability of variable $i$ is obtained by summing over all subsets of parameters including parameter $i$, which yields the total Sobol index for parameter $i$,
\begin{equation}
S_{\lbrace i \}} = \sum_{i \in I} S_{i}. 
\end{equation}
In this work, the Sobol decomposition terms are computed from a generalized polynomial chaos expansion in Legendre polynomials~(\cite{Sudret:Article:2008}), where the expansion coefficients are obtained from sparse quadrature rules using the Smolyak algorithm~(\cite{Smolyak:Article:1963}). This quadrature rule is very sparse but assumes high regularity on the quantity of interest as a function of the input parameters.

\end{document}